\documentclass[trackchanges, twocolumn]{aastex701} %  older version 
\usepackage{multirow, amsmath}  % Allows multirow cells
\usepackage{appendix}
\usepackage{graphicx, subfigure}
\usepackage{acronym}
\usepackage{hyperref}

\DeclareMathOperator{\erf}{erf}

\begin{document}

\acrodef{EMRI}[EMRI]{extreme mass-ratio inspiral}
\acrodef{CO}[CO]{compact object}
\acrodef{LISA}[\textit{LISA}]{\textit{Laser Interferometer Space Antenna}}
\acrodef{GW}[GW]{gravitational wave}
\acrodef{SNR}[SNR]{signal-to-noise ratio}
\acrodef{MBH}[MBH]{massive black hole}
\acrodef{sBH}[sBH]{stellar-mass black hole}
\acrodef{AGN}[AGN]{active galactic nucleus}
\acrodefplural{AGN}[AGNs]{active galactic nuclei}
\acrodef{NSC}[NSC]{nuclear stellar cluster}
\acrodef{MLP}[MLP]{\emph{multi-layer perceptron}}
\acrodef{CI}[CI]{credible interval}
\acrodef{ML}[ML]{machine-learning}
\acrodef{PPD}[PPD]{posterior predictive distribution}

\title{Revealing massive black hole astrophysics: The potential of hierarchical inference with extreme mass-ratio inspiral observations}
\shorttitle{Revealing Massive Black Hole Astrophysics}

\author[orcid=0000-0003-4881-1067,gname='Shashwat', sname='Singh']{Shashwat Singh}
\affiliation{Institute for Gravitational Research, University of Glasgow, Kelvin Building, University Ave., Glasgow, G12 8QQ, United Kingdom}
\email[show]{s.singh.3@research.gla.ac.uk}  

\author[orcid=0000-0002-2728-9612, gname='Christian E. A.', sname='Chapman-Bird']{Christian E. A. Chapman-Bird} 
\affiliation{Institute for Gravitational Wave Astronomy \& School of Physics and Astronomy, University of Birmingham, Edgbaston, Birmingham B15 2TT, United Kingdom}
\email{c.chapman-bird@bham.ac.uk}

\author[orcid=0000-0003-3870-7215, gname='Christopher P. L.', sname='Berry']{Christopher P. L. Berry}
\affiliation{Institute for Gravitational Research, University of Glasgow, Kelvin Building, University Ave., Glasgow, G12 8QQ, United Kingdom}
\email{christopher.berry.2@glasgow.ac.uk}

\author[orcid=0000-0002-6508-0713, gname='John', sname='Veitch']{John Veitch}
\affiliation{Institute for Gravitational Research, University of Glasgow, Kelvin Building, University Ave., Glasgow, G12 8QQ, United Kingdom}
\email{john.veitch@glasgow.ac.uk}

\begin{abstract}
Gravitational waves from extreme mass-ratio inspirals (EMRIs) will enable sub-percent measurements of massive black hole parameters and provide access to the demographics of compact objects in galactic nuclei.
During the \textit{LISA} mission, multiple EMRIs are expected to be detected, allowing statistical studies of massive black hole populations and their formation pathways.
We perform hierarchical Bayesian inference on simulated EMRI catalogues to assess how well \textit{LISA} could constrain the astrophysical population using parametrised population models.
We test our inference framework on a variety of populations, including heterogeneous and homogeneous mixtures of parametrised subpopulations, and scenarios in which the assumed model is deliberately misspecified.
Our results show that population parameters governing distributions with sharp features can be tightly constrained.
Mixed populations can be disentangled with as few as $\sim20$ detections, and even with model misspecification, the inference retains sensitivity to key population features.
These results demonstrate the capabilities and limitations of EMRI population inference, providing guidance for constructing realistic astrophysical population models for \textit{LISA} analysis.
\end{abstract}


\section{Introduction}
\label{sec:Introduction}

% problem of the \ac{MBH} population
Observations reveal a correlation between the properties of \acp{MBH} and their host galaxies~\citep{Richstone:1998ky, Ferrarese:2000se, Kormendy:2013dxa}.
Despite this apparent ubiquity of their co-evolution, the origins and growth of \acp{MBH} remain poorly understood~\citep{Volonteri:2021sfo, Alexander:2025rtn}.
Although new observations from \textit{JWST} shed light on \acp{MBH} accretion in the form of \aclp{AGN}~\citep[\acsp{AGN};][]{2023ApJ...957L...7K, 2025arXiv251010772Z, 2025MNRAS.544..211R, 2023ApJ...959...39H}, the vast majority of \acp{MBH} are electromagnetically quiescent.
The \ac{LISA} mission will help to fill this gap; it will detect \acp{GW} from the merger of black holes across a broad spectrum of masses and redshifts, complementing current electromagnetic observations.

In low-to-intermediate mass galaxies ($\sim10^8$--$10^{12}~M_\odot$), \acp{MBH} are often found co-existing with dense \acp{NSC}, which provide a reservoir for \acp{CO} such as \acp{sBH}, white dwarfs and neutron stars~\citep{Graham:2009un, 2016MNRAS.457.2122G, Neumayer:2020gno}. 
The dynamical evolution of such \acp{CO} in the \ac{NSC} is driven by the gravitational potential of the \ac{MBH} and its environment.
A small fraction of \acp{CO} can become gravitationally bound to the \ac{MBH}, forming a highly asymmetric-mass binary~\citep{Amaro-Seoane:2012lgq}. 
Such a binary system is known as an \ac{EMRI}.
As the \ac{CO} gradually loses energy and angular momentum through the emission of \acp{GW}~\citep{Amaro-Seoane:2007osp, Berry:2019wgg}, the orbit shrinks slowly.
Due to this mass ratio of the binary, an \ac{EMRI} can spend $\sim10^5$--$10^6$ orbital cycles in the millihertz band ($\sim10^{-4}$--$10^{-1}~\mathrm{Hz}$), where \ac{LISA} is most sensitive~\citep{LISA:2024hlh}.
The long-lived and information-rich \ac{GW} signals enable constraints on the \ac{MBH} and \ac{CO} masses, \ac{MBH} spin, and orbital eccentricity and inclination of the orbital plane within $\sim0.001\%$ accuracy~\citep{Barack:2003fp, Babak:2017tow, Chapman-Bird:2025xtd}.
These sub-percent precision measurements of \ac{EMRI} parameters will offer a unique opportunity to explore the \ac{MBH} population~\citep{Gair:2008bx, Babak:2017tow}.

Several \ac{EMRI} formation mechanisms have been proposed depending on the \ac{MBH} environment.
In dense \acp{NSC}, \acp{CO} can be scattered onto high-eccentricity and low-angular-momentum orbits due to gravitational interactions with the stellar background, leading to the formation of a binary with the \ac{MBH}~\citep{1987gady.bookB}.
This mechanism of \ac{EMRI} formation is known as the loss-cone channel~\citep{Lightman:1977zz, Alexander:2017rvg}, and it leads to gradual orbital decay via \ac{GW} emission during successive periapsis passages~\citep{Merritt:2013awa}.
Beyond stellar dynamics, the \ac{MBH} environment plays a critical role in regulating \ac{EMRI} formation.
The presence of gas~\citep{Pan:2021ksp, Pan:2021oob, Lyu:2024gnk}, or its absence~\citep{Alexander:2017aln, Cui:2025bgu}, can alter the inclination of the orbital plane and orbital eccentricity of the \ac{CO}, influencing the \ac{EMRI} rate.
For example, \acsu{AGN} discs could provide an efficient in-situ channel where \acp{CO} form, migrate and inspiral within the disc under the influence of gas drag and disc torques~\citep{2007MNRAS.374..515L, Pan:2021ksp, Speri:2022upm}.
Additional astrophysical factors, such as starbursts~\citep{2007MNRAS.374..515L} or supernova explosions~\citep{Bortolas:2017moe, Bortolas:2019sif}, could further enhance \ac{EMRI} formation rates.
\acp{EMRI} could also form from the tidal capture of one component of a stellar-mass binary~\citep{ColemanMiller:2005rm, Chen:2018axp, Raveh:2020jxg}. 
In such systems, one of the component \ac{CO} could be ejected while the other remains bound to the \ac{MBH}, forming an \ac{EMRI}.
\ac{MBH} binaries could also induce \ac{EMRI} formation \citep{Mazzolari:2022cho}, where the \acp{CO} act as mediators to bring \ac{MBH} binaries to parsec-scales~\citep[Sec.\ 2.2]{LISA:2022yao}.
Hence, given the variety and complexity of the astrophysical processes influencing \ac{EMRI} formation, and the current absence of direct observational constraints, these astrophysical processes and their relative contribution to the total \ac{EMRI} rate remain poorly constrained~\citep{LISA:2022yao}.

By combining multiple \ac{EMRI} observations, it may be possible to constrain both the astrophysical processes governing \ac{EMRI} formation and their relative contributions to the overall population~\citep{Gair:2010yu, Chapman-Bird:2022tvu}.
To extract the astrophysical parameters governing the various astrophysical processes, we need to perform a population-level (\emph{hierarchical}) inference, accounting for selection effects and the uncertainties in individual \ac{EMRI} parameter measurements~\citep{Mandel:2018mve, Vitale:2020aaz}.

Selection effects are important because the detected \ac{EMRI} population is not an accurate representation of the underlying astrophysical population.
Detection efficiency varies strongly across \ac{EMRI} parameter space and is primarily dictated by the \ac{SNR}.
Hence, population inference must explicitly account for these biases to avoid incorrect inference.
In the absence of any standardised \ac{EMRI} detection pipeline, the detectability is commonly modelled as a function of \ac{SNR}~\citep{Gair:2004iv, Gair:2010yu, Babak:2017tow}, equivalent to defining an \ac{SNR}-limited catalogue.
An \ac{SNR}-based selection function quantifies the probability of detecting an event from a given population model, hence factoring in the undetected events~\citep{Mandel:2018mve, Vitale:2020aaz}.
Accurately evaluating this selection function, however, presents a challenge.
It requires computing a large ensemble of \ac{EMRI} \acp{SNR} for a given population model, which in turn requires repeated waveform generation.
This step is a critical computational bottleneck in population inference.
Even with accelerated computation on graphics processing units~\citep{Chapman-Bird:2025xtd}, generating a single 6-year waveform takes a couple of seconds, making direct evaluation of the selection function computationally prohibitive.
To overcome this challenge, \ac{ML}-based emulation, pioneered for \acp{EMRI} in \citet{Chapman-Bird:2022tvu} and further extended in \citet{Singh:2025pzh}, has been developed to address this problem.
Incorporating such \ac{ML}-based emulators into the inference framework allows one to account for undetected events, leading to accurate constraints on the underlying population.

In this work, we apply this hierarchical inference framework to phenomenological population models representing the outcome from various formation channels.
We begin by outlining our framework in Sec.~\ref{sec:Bayesian_Hierarchical_Inference}.
We then discuss our phenomenological population models in Sec.~\ref {sec:population_distribution}, and present results from hierarchical inference in Sec.~\ref {sec:population_inference}, demonstrating the potential for constraining population parameters in a variety of scenarios.
Finally, we conclude in Sec~\ref{sec:conclusion} discussing the scientific potential of \ac{LISA} observations.

\section{Bayesian Hierarchical Inference}
\label{sec:Bayesian_Hierarchical_Inference}

We describe the astrophysical distribution of \acp{EMRI} with a population model $p_{\mathrm{pop}}(\boldsymbol{\theta}|\boldsymbol{\Lambda})$, which gives the probability for single-event parameters $\boldsymbol{\theta}$  and is parametrised by the \emph{hyperparameters} (population-level parameters) $\boldsymbol{\Lambda}$.
Hierarchical Bayesian inference combines multiple observations to infer \(\boldsymbol{\Lambda}\), consistently accounting for the uncertainties arising from single-event parameter estimation.
The goal is to estimate these hyperparameters using a catalogue of \(N_{\mathrm{det}}\) detected \acp{EMRI} from the \ac{LISA} data stream \(\boldsymbol{d}\).

Using Bayes' theorem, we can express the \emph{hyperposterior} (population-level posterior) as 
\begin{equation}
\label{Hierarchical_Inference_eq}
p(\boldsymbol{\Lambda}|\boldsymbol{d}) = \frac{\pi(\boldsymbol{\Lambda})\mathcal{L}(\boldsymbol{d}|\boldsymbol{\Lambda})}{\mathcal{Z}(\boldsymbol{d})},
\end{equation}
where in the numerator, the first term $\pi(\boldsymbol{\Lambda})$ is the \emph{hyperprior} for the population parameters, and the second term is the hyperlikelihood $\mathcal{L}$ for our set of observed data $\boldsymbol{d}$, and the denominator is the \emph{hyperevidence} $\mathcal{Z}$.

% ---------------- POPULATION DESCRIPTION ----------------
\begin{table*}
\centering
\caption{Phenomenological population models considered.
Models A and B are shown explicitly, while mixed models ($\mathrm{A^{(1)}}+\mathrm{A^{(2)}}$, $\mathrm{A}+\mathrm{B}$, $\mathrm{B^{(1)}}+\mathrm{B^{(2)}}$) share the same functional form with a fixed branching fraction of $w = 0.7$.
Each model corresponds to specific functional choices and hyperparameters, as detailed in Table~\ref{tab:generic_form}.
}
\label{tab:models}
\begin{tabular}{ccccc}
\hline\hline
% CHECK JOURNAL STYLE
Parameter ${\theta}_j$ & Description & Model A & Model B & Prior \\
\hline
$\log_{10}(M/M_\odot)$ & Source-frame log \ac{MBH} mass & Schechter & Power-law & [$5$, $10$]\\
$\log_{10}(\mu/M_\odot)$ & Source-frame log \ac{CO} mass & Skew-normal & Power-law & [$1$, $2$]\\
$a$ & \ac{MBH} spin & Beta & Truncated-Normal & [$0.1$, $0.7$]\\
$e_0$ & Initial eccentricity & Beta & Uniform & [$0.1$, $0.7$]\\
$p_0$ & Initial semi-latus rectum & - & - & $p_0(T_\mathrm{plunge})$ \\
$\iota$ & Inclination angle & Uniform & Uniform & [$\arccos(-1)$, $\arccos(1)$]\\
$\theta_\mathrm{S}$ & Polar sky location & Uniform & Uniform & ($0$, $\pi$) \\
$\phi_\mathrm{S}$ & Azimuthal sky location & Uniform & Uniform & ($0$, $2\pi$) \\
$\theta_\mathrm{K}$ & Polar spin orientation & Uniform & Uniform & ($0$, $\pi$) \\
$\phi_\mathrm{K}$ & Azimuthal spin orientation & Uniform & Uniform & ($0$, $2\pi$) \\
$d_\mathrm{L}$ (Gpc) & Luminosity distance & Uniform in $\mathcal{V}_\mathrm{c}$ & Uniform in $\mathcal{V}_\mathrm{c}$ & ($0$, $10$] \\
$\Phi_{\phi,0}$, $\Phi_{\theta,0}$, $\Phi_{r,0}$ & Initial orbital phases & Uniform & Uniform & ($0$, $2\pi$) \\
$T_\mathrm{plunge}$ (yr) & Plunge time since observation start & Uniform & Uniform & [$1$, $6$] \\
\hline
\end{tabular}
\end{table*}

Each observed event $i\in[0,N_{\mathrm{det}}]$ is described by source parameters $\boldsymbol{\theta}_i$ (Table~\ref{tab:models}). 
Under the assumption that detections are independent given the population model $p_{\mathrm{pop}}(\boldsymbol{\theta}|\boldsymbol{\Lambda})$, the \emph{hyperlikelihood}~\citep{Mandel:2018mve, Vitale:2020aaz} can be expressed by marginalising over $\boldsymbol{\theta}_i$ as
\begin{multline}
\label{Hierarchical_Inference_eq_likelihood}
    \mathcal{L} = \prod_{i=1}^{N_{\mathrm{det}}}\frac{\int~\mathrm{d}\boldsymbol{\theta}_i~p(\boldsymbol{d}|\boldsymbol{\theta}_i) p_\mathrm{pop}(\boldsymbol{\theta}_i|\boldsymbol{\Lambda})}{\alpha(\boldsymbol{\Lambda})}  \times \\
    e^{-\mathcal{R\alpha(\boldsymbol{\Lambda})}}[\mathcal{R}\alpha(\boldsymbol{\Lambda})]^{N_{\mathrm{det}}}.
\end{multline}
Here, the product accounts for the marginalised likelihood of each detected event from the chosen population model $p_\mathrm{pop}(\boldsymbol{\theta}|\boldsymbol{\Lambda})$ normalised by the selection function $\alpha(\boldsymbol{\Lambda})$.
The second term accounts for the probability of detecting $N_\mathrm{det}$ events assuming a Poisson process with rate $\mathcal{R}\alpha$, where $\mathcal{R}$ is the rate of \acp{EMRI}. 

The selection function, $\alpha(\boldsymbol{\Lambda})$, is calculated by integrating the probability of detecting a system with given parameters $P_{\mathrm{det}}(\boldsymbol{\theta})$ over the population~\citep{Mandel:2018mve, Vitale:2020aaz}, 
\begin{equation}
\label{sf_expansion}
    \alpha(\boldsymbol{\Lambda}) = \int \mathrm{d}\boldsymbol{\theta} ~ {P}_{\mathrm{det}}(\boldsymbol{\theta})p_\mathrm{pop}(\boldsymbol{\theta}|\boldsymbol{\Lambda}).
\end{equation}
This quantity represents the fraction of the underlying population that is detectable and enables the population-level inference without bias.

Following~\citet{Chapman-Bird:2022tvu}, we treat detectability solely as a function of \ac{SNR} as
\begin{equation}
    P\mathrm{_{det}} = \mathcal{H}(\rho_n-\rho_\mathrm{t}),
\end{equation}
where $\rho_n$ is a noise-realised \ac{SNR}, $\mathcal{H}$ is the Heaviside step function and $\rho_\mathrm{t}$ is a chosen threshold \ac{SNR}.
We obtain $\rho_n^2$ by drawing a sample from a non-central $\chi^2$ distribution with two degrees of freedom and non-centrality parameter $\rho_{\mathrm{opt}}^2$~\citep[Chapter 7]{Maggiore:2007ulw}, where $\rho_{\mathrm{opt}}$ is the optimal \ac{SNR}.

We approximate the selection function in Eq.~\eqref{sf_expansion} by evaluating it as a Monte Carlo sum,
\begin{equation}
    \label{eq:selection_function}
    \alpha(\boldsymbol{\Lambda}) \approx \frac{1}{N_\mathrm{t}}\sum_{k=0}^{N_\mathrm{t}}{\overline{P_\mathrm{det}}}(\boldsymbol{\theta}_k),
\end{equation}
where $\{\boldsymbol{\theta}_k\}$ are sampled from the population distribution $p_\mathrm{pop}(\boldsymbol{\theta}|\boldsymbol{\Lambda})$ and $\overline{P_\mathrm{det}}=1-P(\rho_n>\rho_\mathrm{t}|\rho_\mathrm{opt})$ and $N_\mathrm{t}$ is the number of drawn samples.
To achieve a percent-level accuracy, one needs to compute $P_\mathrm{{det}}(\boldsymbol{\theta}_k)$ $\sim10^5$--$10^6$ times, which is prohibitively expensive considering that $\alpha(\boldsymbol{{\Lambda}})$ must be evaluated a similar number of times during a typical population inference.

To overcome this bottleneck in the selection function, we use \texttt{poplar}~\citep{chapman_bird_2025_17897846}, a developed in \citet{Chapman-Bird:2022tvu} and further extended in \citet{Singh:2025pzh}.
\citet{Chapman-Bird:2022tvu} employed two \ac{MLP} neural networks~\cite[Sec.\ 8.2.1]{Acquaviva:2023gkb} to emulate (i) the \ac{SNR} and (ii) the selection function.
\citet{Singh:2025pzh} further extended and tested the framework, allowing for sources that may or may not plunge, and accounting for random plunges within the finite \ac{LISA} observation window.
Following this approach, we train the selection function \acp{MLP} to map from the astrophysical parameters $\boldsymbol{\Lambda}$ of our phenomenological population model to the selection function $\alpha$.
This framework provides orders-of-magnitude speed-ups in direct evaluations while enabling unbiased estimation of population parameters.

To estimate the hyperposterior $p(\boldsymbol{\Lambda}|\boldsymbol{d})$ in Eq.~\eqref{Hierarchical_Inference_eq_likelihood}, we also need to evaluate the event-level likelihood for the $i$-th event $p(\boldsymbol{d}|\boldsymbol{\theta}_i)$.
The \ac{EMRI} parameters $\boldsymbol{\theta}_i$ can be extracted from the \ac{LISA} data $\boldsymbol{d}$ using event-level parameter-estimation methods.
This requires generating a large ensemble of waveform realisations, which is computationally challenging for the case of \acp{EMRI}.
For computational efficiency, we perform a Fisher matrix analysis where the likelihood is approximated as a multivariate Gaussian~\citep[Chapter 7]{Maggiore:2007ulw}.
This approximation is valid in the high-\ac{SNR} regime, where the likelihood is well-approximated by a Gaussian around the true parameters \citep{Cutler:1994ys, Vallisneri:2007ev}. 
To account for the influence of noise, we randomly offset the maximum likelihood point from the true value, self-consistently using a distribution derived from the Fisher matrix evaluated at the true values \citep{Stevenson:2017dlk, Chapman-Bird:2022tvu}.

We impose an \ac{SNR} threshold $\rho_\mathrm{t}\geq20$ motivated by previous studies on achievable detection thresholds for semi-coherent searches~\citep{MockLISADataChallengeTaskForce:2009wir}.
While this choice is not driven by Fisher matrix requirements, Fisher-based estimates remain accurate at this threshold.
We calculate Fisher matrices using the \texttt{StableEMRIFisher} package~\citep{kejriwal_2024_sef}.
We used the \texttt{Pn5AAK} waveform model for our analysis, implemented in the \texttt{FastEMRIWaveforms} package~\citep{Chua:2020stf, Katz:2021yft, Speri:2023jte, Chapman-Bird:2025xtd}.
Following \citet{Singh:2025pzh}, we adopt an observation window of four years where \ac{EMRI} source may or may not plunge with random plunges mimicking realistic conditions.
The waveforms are then modulated by the \ac{LISA} response function using the first-generation time-delay interferometry configuration with equal arm lengths.
While the choice of time-delay interferometry generation does not significantly impact our results, this setup demonstrates that realistic sky averaging can be naturally incorporated into our framework.
We retain the $A$ and $E$ channels, as the $T$ channel carries negligible strain content in this configuration, and compute the detector response using the \texttt{FastLISAResponse} module of the \texttt{lisa-on-gpu} package~\citep{Katz:2022yqe}.
Having calculated all the necessary quantities to compute the hyperlikelihood, we employ nested sampling using \texttt{nessai}~\citep{nessai, Williams:2021qyt} to explore the hyperparameter space efficiently.

\section{Population Distribution}
\label{sec:population_distribution}

% ---------------- GENERIC TABLE FUNCTIONAL FORM ----------------
\begin{table*}
    \centering
    \caption{Parametric forms used to model the underlying population distributions in this work.
    Each parameter ${\theta}_j \in \boldsymbol{\theta}$ is sampled from a distribution $p_{\mathrm{pop}}({\theta}_j|\boldsymbol{\Lambda})$, characterized by hyperparameters $\boldsymbol{\Lambda}$.}
    \label{tab:generic_form}
    \begin{tabular}{lcc}
        \hline\hline
        Distribution & Functional form $p_{\mathrm{pop}}({\theta}_j|\boldsymbol{\Lambda})$ & Hyperparameters $\boldsymbol{\Lambda}$ \\
        \hline
        Beta & ${\theta}_j^{\alpha-1}(1-{\theta}_j)^{\beta-1}$ & $\alpha$, $\beta$ \\
        Truncated-normal & $\exp[-({\theta}_j-\mu)^2/2\sigma^2] $ & $\mu$, $\sigma$ \\
        Power-law & ${\theta}_j^\lambda$ & $\lambda$ \\
        Schechter & $x_c^{-1}\left({\theta}_j/x_c\right)^{\zeta}\exp\left(-{\theta}_j\zeta/x_c\right)$ & $x_c$ \\
        Skew-normal & $\exp(-{\theta}_j^2/2)~[1+\erf(\gamma{\theta}_j/\sqrt{2})]$ & $\gamma$ \\
        \hline
    \end{tabular}
\end{table*}

We adopt a phenomenological approach to model the \ac{EMRI} population $p_{\mathrm{pop}}(\boldsymbol{\theta}|\boldsymbol{\Lambda})$, parametrising the logarithmic \ac{MBH} ($\log_{10}(M/M_\odot)$; Sec.~\ref{subsec:MBH_mass_spectrum}) and \ac{CO} mass ($\log_{10}(\mu/M_\odot)$; Sec.~\ref{subsec:CO_mass_spectrum}); \ac{MBH} spin ($a$; Sec~\ref{subsec:spin_MBH}) and the initial eccentricity, which we defined at 6 years before the plunge ($e_0$; Sec~\ref{subsec:EMRI_eccentricity}).

We consider two representative \ac{EMRI} population models, A and B, described in Table~\ref{tab:models}, which differ in complexity, and combine them to form mixed models: $\mathrm{A^{(1)}}+\mathrm{A^{(2)}}$, $\mathrm{A}+\mathrm{B}$, and $\mathrm{B^{(1)}}+\mathrm{B^{(2)}}$.
Model A represents hierarchical and accretion-driven \ac{MBH} growth and evolution motivated from literature~\citep{Wang:2025dcd, Sijacki:2014yfa, 2021MNRAS.503.1940H, Somerville:2008bx}, accompanied by mass segregation~\citep{1977ApJ...216..883B, Alexander:2017rvg} in galactic nuclei. 
Additionally, spins are consistent with efficient spin-up or spin-down through prolonged accretion and mergers~\citep{1974ApJ...191..507T, King:2006uu}.
The eccentricity distribution reflects expectations for gradual circularisation in gas-rich nuclei~\citep{Pan:2021oob} or high eccentricities for other channels.
In contrast, Model B represents a simplified scenario in which \ac{MBH} growth is modelled using a power-law distribution for the \ac{MBH} and \ac{CO} masses, with the slope parameter governing the relative abundances of these masses~\citep{Babak:2017tow, LIGOScientific:2020kqk}.
\ac{MBH} spin in Model B follows a narrow normal distribution signifying the overabundance of the dominant growth scenario, and we model eccentricities with a uniform distribution. 
These two models illustrate contrasting assumptions about \ac{EMRI} populations, and serve as a foundation for exploring mixed formation scenarios and assessing the impact of model complexity on population predictions.

In the case of mixed models, we construct synthetic populations, such as $\mathrm{A}+\mathrm{B}$, by combining two distinct population models, thereby allowing us to explore the inference framework's ability to disentangle heterogeneous formation channels.
Similarly, we create homogeneous mixed models ($\mathrm{A^{(1)}}+\mathrm{A^{(2)}}$ and $\mathrm{B^{(1)}}+\mathrm{B^{(2)}}$) by drawing two subpopulations from the same underlying model type but with contrasting hyperparameter values.
These mixed models are constructed to test whether the framework can identify subpopulations within a single formation scenario with distinct hyperparameters or distinguish between completely different channels.
For all analyses of mixed-population models, we fix the branching fraction at $w=0.7$ for consistency.
The mixed population is defined as 
\begin{multline*}
    p_\mathrm{pop}^{X+Y}(\boldsymbol{\theta}|\boldsymbol{\Lambda}) = wp_\mathrm{pop}^{X}(\boldsymbol{\theta}|\boldsymbol{\Lambda}^{X}) + (1-w)p_\mathrm{pop}^{Y}(\boldsymbol{\theta}|\boldsymbol{\Lambda}^{Y}),
\end{multline*}
where $X, Y \in \mathrm{\{A, B\}}$, and $\boldsymbol{\Lambda}^X, \boldsymbol{\Lambda}^Y$ refer to the hyperparameters of the respective specific submodels.
These mixed models enable us to test astrophysically motivated scenarios that span different assumptions about \ac{MBH} growth, \ac{CO} mass spectra, and orbital dynamics.
Representative functional forms are listed in Table~\ref{tab:generic_form} and the hyperprior ranges in Table~\ref{tab:hyperprior_values}.

\subsection{MBH mass spectrum}
\label{subsec:MBH_mass_spectrum}

The mass spectrum of \acp{MBH} is shaped by the distribution of initial seed masses and their subsequent growth through accretion and mergers~\citep{Volonteri:2021sfo}.
Multiple seed formation channels populate the early Universe with characteristic initial masses.
Population III stars~\citep{Klessen:2018fep} in the early metal-free Universe set the initial black hole mass at $\sim 10^2$--$10^3~M_\odot$~\citep{Heger:2001cd}.
In massive dark matter halos ($\sim10^8~M_\odot$), conditions allow for direct collapse of gas leading to the formation of massive seeds of $\sim10^5$--$10^7~M_\odot$~\citep{Begelman:2006db}.
Dense stellar clusters may experience runaway collisions of massive stars, potentially producing very massive stars that collapse into intermediate-mass black holes~\citep[$\sim 2\times10^2$--$10^3~M_\odot$;][]{PortegiesZwart:2004ggg, Mapelli:2016vca}.
These seeds subsequently grow through gas accretion and mergers, producing a mass distribution shaped by hierarchical galaxy assembly and feedback-regulated accretion.

To capture this diversity in our population models, we consider two alternative parameterisations for the \ac{MBH} mass distribution (Table~\ref{tab:models}).
For we adopt a modified Schechter-like distribution~\citep{1976ApJ...203..297S} for Model A and a power-law distribution for Model B~\citep{Babak:2017tow, Barausse:2020mdt, LIGOScientific:2020kqk, 2021MNRAS.503.1940H}.
These two distributions allow exploring the performance of the hierarchical inference framework given the presence or absence of structure in the \ac{MBH} mass spectrum.

The Schechter distribution is widely used to describe galaxy and halo mass functions in hierarchical structure formation~\citep{1974ApJ...187..425P, 1976ApJ...203..297S}.
Hierarchical galaxy assembly and feedback-regulated accretion can produce a mass spectrum with a characteristic peak~\citep{Wang:2025dcd, Sijacki:2014yfa}.
Since \acp{MBH} co-evolve with galaxies~\citep{Somerville:2008bx}, we adopt a modified Schechter distribution to model \ac{MBH} mass distribution for Model A.
In the inference, we vary the hyperparameter $x_c$, which represents the peak in \ac{MBH} mass spectrum and fix the slope $\zeta=7$ (Table~\ref{tab:generic_form}).
The prior range for $x_c$ (Table~\ref{tab:hyperprior_values}) and the functional form is motivated by numerical simulations~\citep{Wang:2025dcd, Sijacki:2014yfa, 2021MNRAS.503.1940H} and semi-analytic models of \ac{MBH}--galaxy co-evolution~\citep{Somerville:2008bx}.

In contrast, the slope parameter $\lambda_M$ used for Model B reflects the relative abundance of light versus heavy \acp{MBH}.
A positive slope $\lambda_M > 0$ favours heavier \acp{MBH}, while $\lambda_M < 0$ suggest abundance of lighter \acp{MBH}. 
We explore $\lambda_M$ within a range (Table~\ref{tab:hyperprior_values}) consistent with both cosmological simulations~\citep{2021MNRAS.503.1940H, Barausse:2012fy} and empirical fits to \ac{GW} source populations~\citep{Babak:2017tow, LIGOScientific:2020kqk}.

By adopting these parametrisations, our population models remain physically motivated and robust across plausible formation channels.
Additional processes such as metal-poor gas inflows~\citep{Lupi:2014vza}, galaxy interactions~\citep{Volonteri:2002vz, Hopkins:2005fb}, feedback~\citep{2012ARA&A..50..455F, DiMatteo:2005ttp} and \ac{GW} kicks~\citep{Baker:2008md} could further modulate growth and alter the mass spectrum, but their inclusion is beyond the scope of our illustrative models.

\begin{table*}
\centering
\caption{
    Hyperprior ranges for the population models considered in this work.
    Each column corresponds to the population models: single-component (A or B) or mixture models ($\mathrm{A^{(1)}}+\mathrm{A^{(2)}}$, $\mathrm{A}+\mathrm{B}$, $\mathrm{B^{(1)}}+\mathrm{B^{(2)}}$).
    For each hyperparameter $\Lambda_j \in \boldsymbol{\Lambda}$, the hyperprior $\pi({\Lambda_j})$ is set as a uniform distribution with the range $\Lambda_j \in [{\Lambda}_{\min}, {\Lambda}_{\max}]$ specified.
    Superscripts $\mathrm{A}$ and $\mathrm{B}$ denote the component population in the mixed population model, while $w$ is the branching fraction.
    A dash (--) indicates that a parameter is not included in the corresponding model.
}
\label{tab:hyperprior_values}
\begin{tabular}{lccccc}
\toprule
Parameters $\boldsymbol{\theta}$ & A & B & $\mathrm{A^{(1)}}+\mathrm{A^{(2)}}$ & $\mathrm{A}+\mathrm{B}$ & $\mathrm{B^{(1)}}+\mathrm{B^{(2)}}$ \\

\hline\hline

% MBH mass ---------------------------------------------------------------
\multirow{2}{*}{MBH mass}
& \multirow{2}{*}{$x_c \in [6.5,~9.5]$}
& \multirow{2}{*}{$\lambda_M \in [-3,~-1.01]$}
& $x_c^{\mathrm{A^{(1)}}} \in [6.5,~9.5]$
& $x_c^{\mathrm{A}} \in [6.5,~9.5]$& $\lambda_M^{\mathrm{A^{(1)}}} \in [-3,~-1.01]$
\\
& & 
& $x_c^{\mathrm{A^{(2)}}} \in [6.5,~9.5]$
& $\lambda_M^{\mathrm{B}} \in [-3,~-1.01]$
& $\lambda_M^{\mathrm{B^{(2)}}} \in [-3,~-1.01]$
\\[0.1em]

\hline

% CO mass ---------------------------------------------------------------
\multirow{2}{*}{CO mass}
& \multirow{2}{*}{$\gamma_\mu \in [-1,~1]$}
& \multirow{2}{*}{$\lambda_\mu \in [-4,~-1.5]$}
& $\gamma_\mu^{\mathrm{A^{(1)}}} \in [-1,~1]$
& $\gamma_\mu^{\mathrm{A}} \in [-1,~1]$
& $\lambda_\mu^{\mathrm{B^{(1)}}} \in [-4,~-1.5]$
\\
& & &
  $\gamma_\mu^{\mathrm{A^{(2)}}} \in [-1,~1]$& $\lambda_\mu^{\mathrm{B}} \in [-4,~-1.5]$
& $\lambda_\mu^{\mathrm{B^{(2)}}} \in [-4,~-1.5]$
\\[0.1em]

\hline

% Spin ------------------------------------------------------------------
\multirow{4}{*}{Spin}
& \multirow{2}{*}{$\alpha_a \in [10,~15]$}
& \multirow{2}{*}{$\mu_a \in [0.1,~0.7]$}
& $\alpha_a^{\mathrm{A^{(1)}}} \in [10,~15]$
& $\alpha_a^{\mathrm{A}} \in [10,~15]$
& $\mu_a^{\mathrm{B^{(1)}}} \in [0.1,~0.7]$
\\

& \multirow{4}{*}{$\beta_a \in [5,~10]$}
& \multirow{4}{*}{$\sigma_a \in [0.001,~0.05]$}
& $\beta_a^{\mathrm{A^{(1)}}} \in [5,~10]$
& $\beta_a^{\mathrm{A}} \in [5,~10]$
& $\sigma_a^{\mathrm{B^{(1)}}} \in [0.001,~0.05]$
\\

& & 
& $\alpha_a^{\mathrm{A^{(2)}}} \in [10,~15]$& $\mu_a^{\mathrm{B}} \in [0.1,~0.7]$
& $\mu_a^{\mathrm{B^{(2)}}} \in [0.1,~0.7]$
\\

& & 
& $\beta_a^{\mathrm{A^{(2)}}} \in [5,~10]$& $\sigma_a^{\mathrm{B}} \in [0.001,~0.05]$
& $\sigma_a^{\mathrm{B^{(2)}}} \in [0.001,~0.05]$
\\[0.1em]

\hline

% Eccentricity ------------------------------------------------------------
\multirow{4}{*}{Eccentricity}
& \multirow{2}{*}{$\alpha_e \in [5,~10]$}
& \multirow{2}{*}{--}
& $\alpha_{e_0}^{\mathrm{A^{(1)}}} \in [5,~10]$
& $\alpha_{e_0}^{\mathrm{A}} \in [5,~10]$
& --
\\

& \multirow{4}{*}{$\beta_e \in [1,~5]$}
& \multirow{4}{*}{--}
& $\beta_{e_0}^{\mathrm{A^{(1)}}} \in [1,~5]$
& $\beta_{e_0}^{\mathrm{A}} \in [1,~5]$
& --
\\

& &
& $\alpha_{e_0}^{\mathrm{A^{(2)}}} \in [5,~10]$& --
& --
\\

& &
& $\beta_{e_0}^{\mathrm{A^{(2)}}} \in [1,~5]$& --
& --
\\[0.1em]

\hline

% Weight ---------------------------------------------------------------
\multirow{1}{*}{Weight}
& --
& --
& $w \in [0.001,~0.999]$& $w \in [0.001,~0.999]$& $w \in [0.001,~0.999]$\\

\hline

\end{tabular}
\end{table*}

\subsection{MBH spin}
\label{subsec:spin_MBH}

The spin of a \ac{MBH} is closely linked to its evolutionary history.
The two principal growth channels are gas accretion and hierarchical mergers, and the spin distribution encodes which process dominates~\citep{Beckmann:2024ezt, Volonteri:2004cf}.
Spin evolution is not monotonic: \acp{MBH} can be spun up or down in an erratic manner, with the spin expected to change repeatedly throughout their lifetime.

To capture the complexity of \ac{MBH} spin distributions arising from the interplay of accretion, mergers, and galaxy properties, we adopt phenomenological population models that are flexible enough to represent a wide range of astrophysical scenarios.
For Model A, we use a beta distribution~\citep{Wysocki:2018mpo}, where the hyperparameters $\alpha_a$ and $\beta_a$ control the shape of the distribution.
Configurations with $\alpha_a < \beta_a$ favour an abundance of low-spin \acp{MBH} expected when stochastic accretion dominates, whereas $\alpha_a > \beta_a$ produces a distribution biased toward high-spin \acp{MBH} consistent with prolonged coherent accretion.
For Model B, we employ a truncated normal distribution (Table~\ref{tab:models}), which captures spins concentrated around a mean value $\mu_a$ with a characteristic spread $\sigma_a$, while enforcing the physical bounds $0 \le a \le 0.998$.
This reflects scenarios where spins cluster around a characteristic value, where one mechanism dominates over the others.
These choices offer flexibility to replicate the diverse spin distributions anticipated from the underlying astrophysical processes.
The prior ranges for these hyperparameters (Table~\ref{tab:hyperprior_values}) are informed by cosmological simulations and semi-analytic models of MBH evolution~\citep{Sesana:2014bea, Izquierdo-Villalba:2020hfk}.

The spin distribution can be influenced by various factors. 
Coherent accretion can efficiently spin up a \ac{MBH}~\citep[to a maximum $a=0.998$;][]{1974ApJ...191..507T}, whereas stochastic, randomly oriented accretion events drive the spin toward moderate values~\citep{King:2006uu, Berti:2008af, Fanidakis:2009ct}.
\ac{MBH} mergers can abruptly alter the spin, typically producing significant reorientations and either spin-up or spin-down depending on the mass ratio and orbital configuration, effectively erasing any memory of the progenitor spins~\citep{Dotti:2009vz}.
The spin magnitude is expected to evolve with redshift~\citep{Beckmann:2024ezt, Volonteri:2004cf, Sala:2023uxf}, although we do not model this explicitly.
At higher redshifts, gas-rich environments spin up the \ac{MBH}, while at lower redshifts the spin evolution diverges depending on \ac{MBH} mass~\citep{Volonteri:2012yn}.
Given these competing effects, the true population shape could exhibit skewness or tails.
For this study, we therefore adopt flexible population models, including a beta distribution that can model strong boundary concentrations and skewness, as well as a truncated normal distribution that captures clustering around a characteristic value.

\subsection{CO mass spectrum}
\label{subsec:CO_mass_spectrum}

\acp{NSC}, which host an abundance of such \acp{CO}, provide the natural reservoir for \ac{EMRI} formation.
However, only a small fraction of these \acp{CO} will form an \ac{EMRI}.

A key mechanism that sustains the supply of \acp{CO} near the \ac{MBH} is mass segregation: a two-body relaxation process where the heavier \ac{CO}, such as \acp{sBH}, sinks towards the galactic nucleus, pushing the lighter objects, such as stars, outward~\citep{1977ApJ...216..883B, Alexander:2017rvg}. 
This affects the mass distribution of stellar \acp{CO} in the \ac{NSC}~\citep{Hopman:2006xn, Aharon:2016kil, Amaro-Seoane:2010dzj}.
Depending upon the abundance of \acp{CO}, mass segregation can fall into one of the two branches: weak~\citep{1977ApJ...216..883B} and strong~\citep{Alexander:2008tq} regimes.
Strong mass segregation follows in the scarcity of heavy \ac{CO}, undergoing scattering with the lighter components before sinking towards the centre due to dynamical friction against the sea of lighter objects, developing a quasi-steady state~\citep{Kaur:2024hfh, Preto:2009kd}.
Conversely, in the weak regime, heavy \acp{CO} are relatively abundant and undergo minimal interactions with the lighter objects, behaving as a single-mass population. 
We model this relative abundance in Model A using a truncated skew-normal distribution.
The hyperparameter $\gamma_\mu$ controls the skewness of the distribution: $\gamma_\mu < 0$ corresponds to a scarcity of heavy \acp{sBH}, reflecting a strong regime, while $\gamma_\mu > 0$ indicates an abundance of more massive \acp{sBH}, as expected in the weak regime.
For comparison, in Model B, we adopt a power-law distribution~\citep{LIGOScientific:2020kqk}.
The slope of the power-law $\lambda_\mu$ governs the relative abundance of \acp{CO}: $\lambda_\mu > 0$ favours more massive \acp{sBH}, while $\lambda_\mu < 0$ suppresses their presence in the \ac{NSC}.

For both the models, we restrict our analysis to \acp{sBH} in the range $10M_\odot$–$100~M_\odot$~\citep{Gair:2004iv}.

\subsection{Initial Eccentricity}
\label{subsec:EMRI_eccentricity}

For the case of \acp{EMRI}, high initial eccentricities ($>0.99$) are expected during the relaxation~\citep{Amaro-Seoane:2007osp, Broggi:2022udp, Mancieri:2024sfy}.
However, much of it will be lost in orbital evolution before the signal enters the \ac{LISA} band~\citep{Peters:1963ux}, broadening the distribution~\citep{Amaro-Seoane:2020zbo}.
Initial eccentricities strongly depend on the formation channel.
\acp{CO} captured through two-body relaxation typically begin on extremely eccentric orbits (e.g., $e_0>0.9$), as they are scattered onto plunging trajectories with small periapsis radii~\citep{Amaro-Seoane:2007osp, Alexander:2017rvg}.
In contrast, objects embedded in \ac{AGN} discs often undergo significant circularisation due to viscous torques and gas dynamical friction, resulting in nearly circular orbits by the time they enter the \ac{LISA} band~\citep{Pan:2021ksp}.
Binaries that are tidally separated (b-\acp{EMRI}) are expected to have low ($e_0<0.01$) to moderate ($e_0 < 0.1$) eccentricities~\citep{ColemanMiller:2005rm}.
Even with effective circularisation, eccentricities may remain measurable, depending on their initial conditions and inspiral timescales.
This eccentricity, hence, becomes a powerful diagnostic: highly eccentric inspirals suggest a dynamical origin, whereas circular EMRIs are more consistent with disc migration scenarios.

To capture the diversity of \ac{EMRI} eccentricities arising from different scenarios, we model the eccentricity distribution with a beta distribution (Table~\ref{tab:models}) for Model A and a uniform distribution for Model B.
For Model A, the hyperparameters $\alpha_{e_0}$ and $\beta_{e_0}$ control the relative weight of high versus low-eccentric systems.
Configurations with $\alpha_{e_0} > \beta_{e_0}$ represent populations dominated by dynamical capture through two-body relaxation, where highly eccentric orbits are typical, whereas $\alpha_{e_0} < \beta_{e_0}$ corresponds to scenarios producing predominantly low eccentricities, such as \acp{EMRI} embedded in \ac{AGN} discs or b-\acp{EMRI} from binary tidal separations.
In contrast, Model B assumes a uniform distribution (Table~\ref{tab:models}), representing the case where no single formation channel dominates, and eccentricities are spread with equal probabilities across the full physical range.

There are several other factors that could influence \ac{EMRI} eccentricities. 
Environmental effects such as stellar scattering, mass segregation, and resonant relaxation~\citep{Gurkan:2007bj} can modify orbital eccentricity before \ac{GW} emission dominates~\citep{Broggi:2022udp}.
Eccentricity evolution is also expected to depend on the redshift and mass of the central MBH~\citep{Amaro-Seoane:2012lgq}.
While these effects are not modelled here, they represent a potential extension for future work.

\subsection{Inclination, Luminosity Distance}
\label{subsec:inclination_Luminosity_dist}

Our baseline assumption is that the inclination angle $\iota_0$ is uniformly distributed in $\cos \iota_0 \in [-1, 1]$, where positive (negative) values correspond to prograde (retrograde) orbits.
While this is a common choice~\citep{Amaro-Seoane:2007osp, Gair:2010yu} for isotropic populations, it is a simplification. 
In realistic environments, inclination distributions can be strongly biased by formation channels and local dynamics.
For example, \acp{EMRI} forming in \ac{AGN} discs are expected to align with the disc plane due to gas torques and migration, favouring low inclinations relative to the disc axis~\citep{Pan:2021ksp, Lyu:2024gnk}.
Similarly, mass segregation in dense stellar clusters can lead to preferential capture of objects on prograde orbits around spinning \acp{MBH}, while resonant relaxation may introduce mild anisotropies~\citep{Hopman:2006xn, Alexander:2008tq, Alexander:2017rvg}.
These effects could produce non-uniform inclination distributions~\citep{Zhang:2025jmm}, particularly in galactic nuclei with significant rotation or disc structure. 
Incorporating such correlations between inclination and environment is beyond the scope of this work, but represents an important extension for future studies.

For luminosity distance, we distribute sources uniformly in comoving volume, up to a luminosity distance of $d_\mathrm{L}=10~\mathrm{Gyr}$, assuming a $\Lambda$--cold dark matter cosmology with $\Omega_M=0.3$ and $\Omega_\Lambda=0.7$.
This neglects any redshift evolution of the source population.
The \ac{EMRI} population is likely to evolve with redshift as both the distribution of \acp{MBH} and \acp{CO} changes across cosmic time.
Therefore, incorporating more realistic redshift distributions~\citep{Gennari:2025nho}, potentially correlated with other parameters, is important for precision modelling. 
We leave studying these potential correlations for future work, as they require more sophisticated population models.

\section{Population Inference}
\label{sec:population_inference}

We conduct a case-by-case study for the population models discussed in Sec.~\ref{sec:population_distribution}.
The default results presented are for $10^4$ injections, but the figures also show results from smaller injection sets ($10^2$ and $10^3$) for comparison. 
The fraction of detected events remains consistent across the number of injections, staying within the range $\sim17$--$21\%$ given the \ac{SNR} threshold of $20$.

First (Sec.~\ref{case_1_inference}), we test our inference framework on single-component populations (Models A and B) to verify its ability to accurately recover injected hyperparameters.
Next (Sec.~\ref{case_2_inference}), we consider mixed populations constructed in two ways: (i) heterogeneous mixtures combining distinct astrophysical models ($\mathrm{A}+\mathrm{B}$), and (ii) homogeneous mixtures using the same functional form but extreme hyperparameter values ($\mathrm{A^{(1)}}+\mathrm{A^{(2)}}$, $\mathrm{B^{(1)}}+\mathrm{B^{(2)}}$).
This mixture of heterogeneous and homogenous population models allows us to investigate whether we can differentiate between contrasting population models.
Finally (Sec.~\ref{case_3_inference}), we test model \emph{misspecification} by inferring populations with an incorrect population model.
We inferred Model $\mathrm{A}+\mathrm{B}$ using Models A and B separately, and similarly, Model A (B) using Model B (A) and $\mathrm{A}+\mathrm{B}$.
These checks enable us to investigate what would happen in a plausible scenario where a parametrised population model does not accurately reflect the underlying astrophysical population.

\subsection{Inference on individual population models ($\mathrm{A}$, $\mathrm{B}$)}
\label{case_1_inference}

\begin{figure*}
    \centering
    \includegraphics[width=18cm, height=10cm]{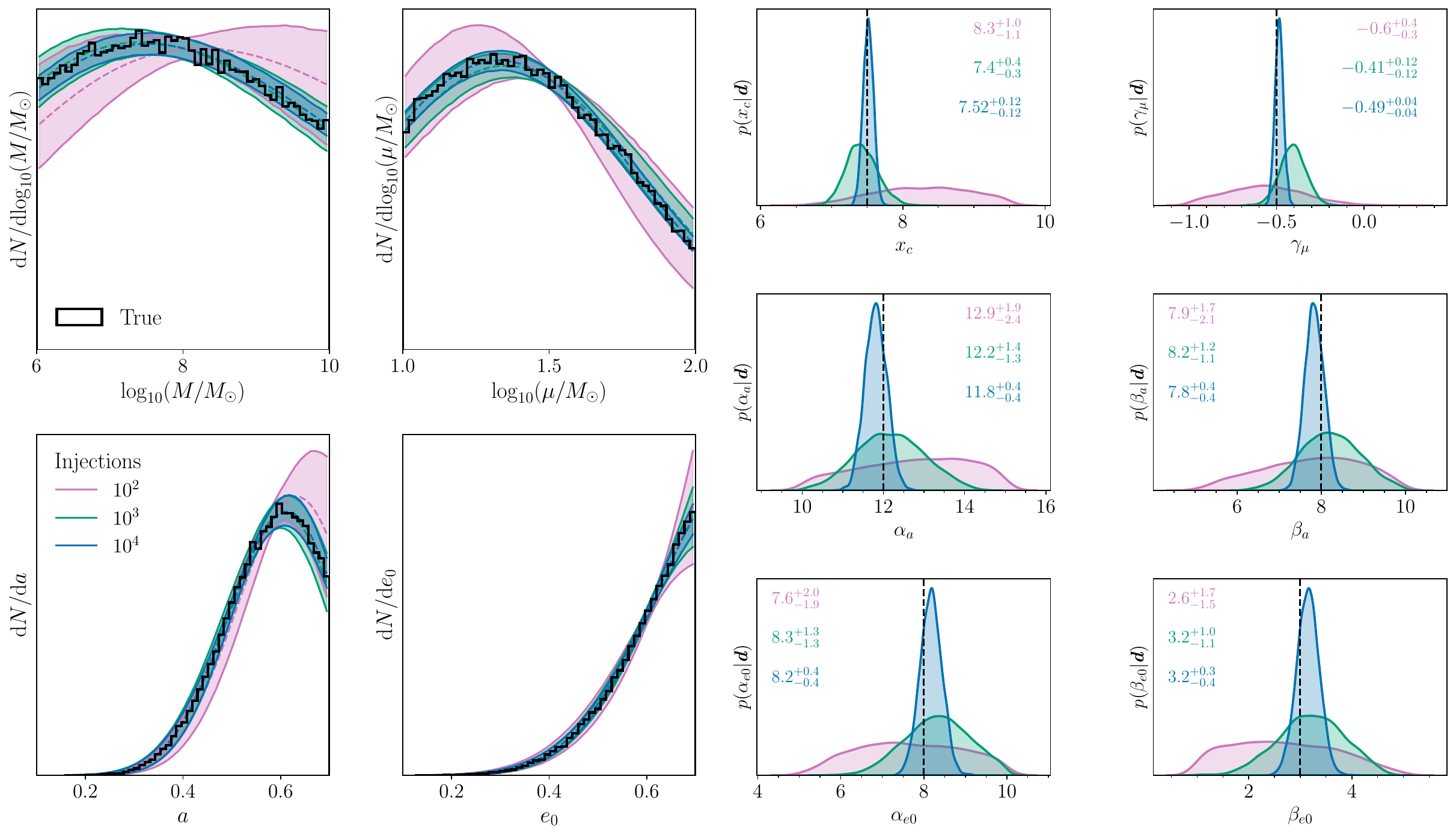}
    \caption{Posterior predictive distributions (\textit{left}) and one-dimensional hyperparameter posteriors (\textit{right}) for Model A.
    In the predictive distribution panels, the coloured solid curves denote the $90$-th percentile and the dashed curves denote the mean.
    The colour scheme is consistent across panels: pink, green, and blue correspond to $10^2$, $10^3$, and $10^4$ \ac{EMRI} injections, respectively.
    For each set of injections $10^2$, $10^3$, and $10^4$ for these simulations we detect $18$, $188$ and $1886$ events, respectively.
    The solid black curve shows the true distribution reconstructed from the injected hyperparameters, whose values are marked by the vertical black dashed lines in the one-dimensional posteriors.
    The predictive distributions display the differential number density $\mathrm{d}N/\mathrm{d}\boldsymbol{\theta}_i$ for $\boldsymbol{\theta}_i \in \{\log_{10}(M/M_\odot), \log_{10}(\mu/M_\odot), a, e_0\}$, with symbols defined in Table~\ref{tab:models}.
    The one-dimensional posteriors summarise the inferred hyperparameters from the hierarchical population analysis.
    Each subplot on the \textit{right} reports the median and associated $90\%$ uncertainties, with colours indicating the number of injected events.
    The horizontal axis spans the relevant portion of the prior.
    }
\label{pop_A}
\end{figure*}

\begin{figure*}
    \centering
    \includegraphics[width=18cm, height=10cm]{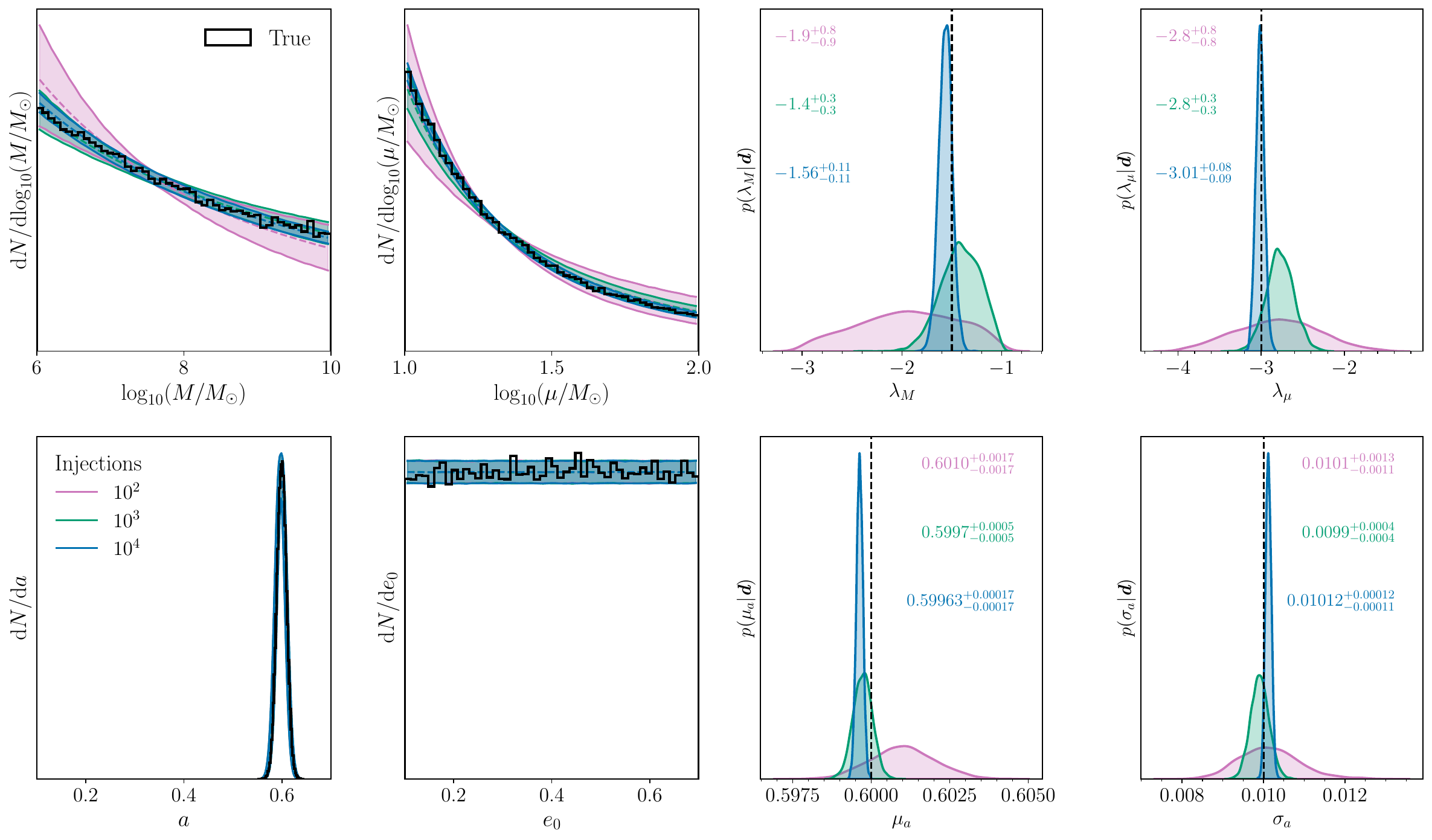}
    \caption{Posterior predictive distributions (\textit{left}) and one-dimensional hyperparameter posteriors (\textit{right}) for Model B.
    The plotting follows the same structure as Figure~\ref{pop_A}, where the only difference is the number of hyperparameters used to model population Model B, where the number of detections is $21$, $206$ and $2080$, for $10^2$, $10^3$, and $10^4$ injections, respectively.
    }
\label{pop_B}
\end{figure*}

To assess the performance and consistency of hierarchical inference, we use the \ac{PPD}.
\acp{PPD} quantifies the distribution of future or unobserved event parameters predicted by the inferred population model. 
It explicitly accounts for uncertainty in the population hyperparameters.

Given a set of observed data $\boldsymbol{d} = \{d_i\}$ and population hyperparameters $\boldsymbol{\Lambda}$, the \ac{PPD} for a new event with parameters $\boldsymbol{\tilde{\theta}}$ is given by
\begin{equation}
p(\tilde{\theta}|\boldsymbol{d}) = \int p(\tilde{\boldsymbol{\theta}}|\boldsymbol{\Lambda})~p(\boldsymbol{\Lambda}|\boldsymbol{d})~\mathrm{d}\boldsymbol{\Lambda},
\end{equation}
where $p(\boldsymbol{\tilde{\theta}}|\boldsymbol{\Lambda})$ is the population model and $p(\boldsymbol{\Lambda}|\boldsymbol{d})$ is the hyperposterior inferred from the observed catalogue.

In Figure~\ref{pop_A} and Figure~\ref{pop_B}, we show the \acp{PPD} and the recovered one-dimensional posterior for Models A and B, respectively.
The injected (true) values typically lie well within the inferred $90\%$ credible intervals, with some fluctuations depending upon the exact realisation of the catalogue of detections and simulated noise for each signal.
As expected, increasing the number of \ac{EMRI} injections significantly tightens the posterior distributions, improving the precision of the recovered hyperparameters.
This confirms the consistency of the inference pipeline when the population model matches the underlying distribution.

\ac{PPD} plots also confirm the same trend.
The uncertainty in the recovered hyperparameters approximately follows the expected $\propto 1/\sqrt{N_\mathrm{det}}$ scaling with the number of detections $N_\mathrm{det}$.
Constraints tighten rapidly between $10^2$ and $10^3$ injections, with gradual improvement for higher injections, and the rate of improvement diminishes with larger sample sizes.
However, even with $10^2$ injections ($18$ detections), we recover well-informed posteriors. 

For Model A, the hyperparameter representing the peak of the \ac{MBH} mass spectrum $x_c$ is recovered within $1.5\%$ uncertainty for $10^4$ injections ($1886$ detections), indicating a tight constraint.
By comparison, the spin hyperparameters $\alpha_a$ and $\beta_a$ and the eccentricity parameters  $\alpha_{e0}$ and $\beta_{e0}$ are recovered with larger fractional errors, though still well constrained.
We recover $\alpha_a$ and $\beta_a$ within $\sim2\%$ and $\sim3\%$, and $\alpha_{e0}$ and $\beta_{e0}$ within $\sim3\%$ and $\sim4\%$.
The relative abundance of \acp{CO} in the \ac{NSC}, represented in Model A as $\gamma_\mu$, is recovered within $\sim8\%$, which is the worst for this population model.
Overall, we observe that for Model A, the parameters directly related to the \ac{MBH}: the peak of the mass spectrum and its shape parameters governing the spin distribution, are recovered with the highest precision across all injection sets.
This highlights that \ac{EMRI} observations are particularly sensitive to the \ac{MBH} mass and the spin.

For Model B, hyperparameters governing the \ac{MBH} spin are well constrained, with sub-percent fractional uncertainty.
This demonstrates that an abundance of a specific range of spin parameters will leave a clear imprint on the \ac{MBH} spin, allowing for sub-percent recovery even for relatively small numbers of detected events ($21$).
The slope of the \ac{MBH} and \ac{CO} mass $\lambda_M$ and $\lambda_\mu$ are recovered within $\sim4\%$ for $10^4$ injections ($2080$ detections).

Comparing the two population models, differences in parameter precision primarily reflect the structure of the underlying distributions.
In Model A, the \ac{MBH} and the \ac{CO} mass spectra exhibit characteristic features: the \ac{MBH} mass distribution peaks at $x_c$, and the \ac{CO} skewness parameter $\gamma_\mu$ follows the injected high abundance of low-mass \acp{sBH}.
This feature enables their recovery with high precision.
By contrast, both in Model B are simple power laws, which lack sharp features and are consequently harder to constrain. 
Furthermore, the \ac{MBH} spin distribution in Model A is described by a beta distribution, which is more flexible than the Gaussian-like spin distribution adopted in Model B and therefore results in larger fractional uncertainties, whereas the narrower Gaussian-like \ac{MBH} spin distribution in Model B allows sub-percent precision.
Hence, hyperparameters that control sharply defined features are consistently inferred with higher accuracy, whereas hyperparameters governing broad distributions yield correspondingly weaker constraints \citep{Chakrabarty:2003kt, Fishbach:2019bbm}.

\subsection{Inference on model mixtures ($\mathrm{A}+\mathrm{B}$, $\mathrm{A^{(1)}}+\mathrm{A^{(2)}}$, $\mathrm{B^{(1)}}+\mathrm{B^{(2)}}$)}
\label{case_2_inference}

\begin{figure*}
    \centering
    \includegraphics[width=17cm, height=13cm]{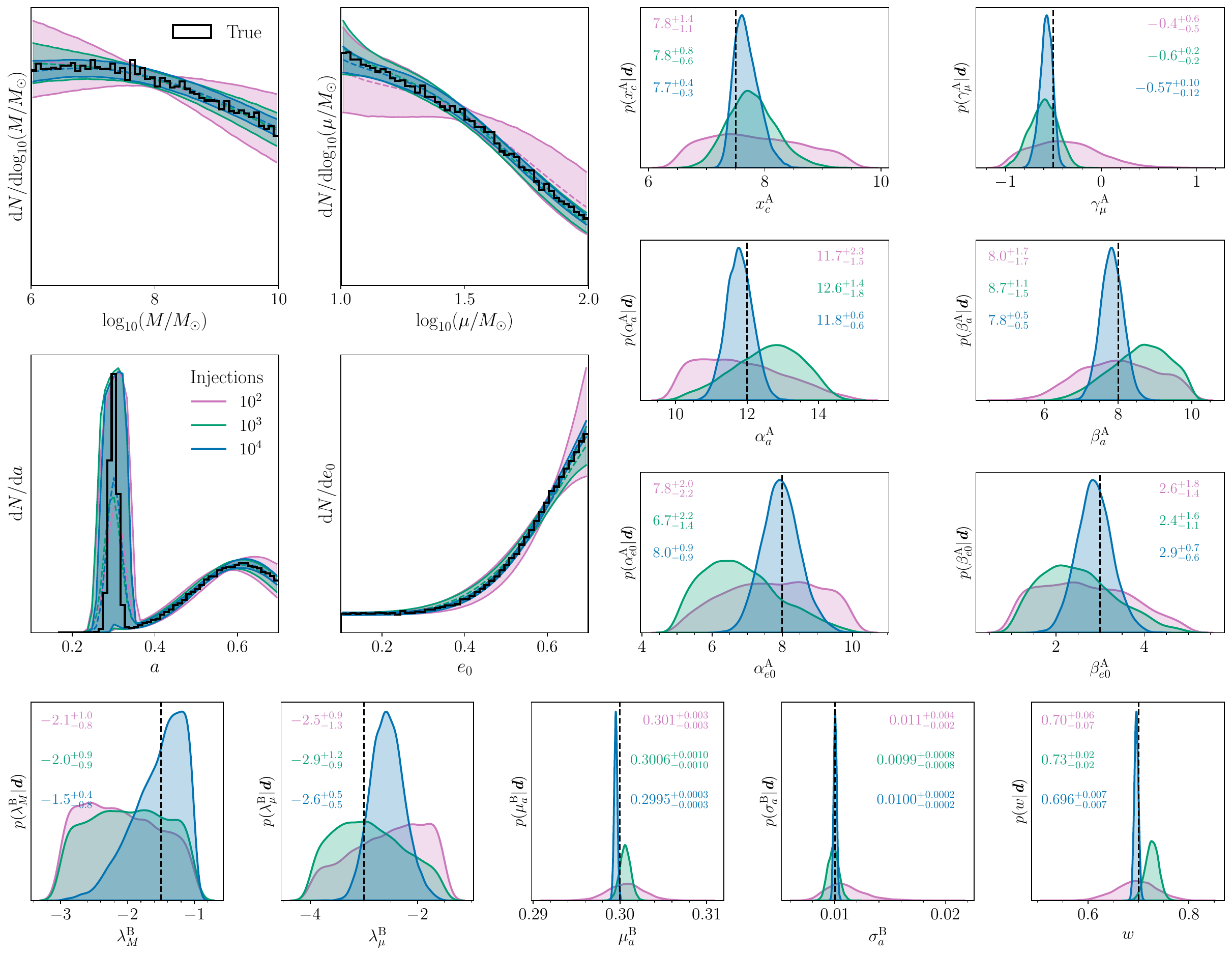}
    \caption{Posterior predictive distributions (\textit{upper left}) and one-dimensional hyperparameter posteriors (\textit{right} and \textit{bottom}) for Model $\mathrm{A}+\mathrm{B}$. 
    The plotting follows the same structure as Figure~\ref{pop_A}, where the only difference is the number of hyperparameters used to model the mixed population, where the number of detections is $19$, $198$ and $1980$, for $10^2$, $10^3$, and $10^4$ injections, respectively.
    Results for other mixed populations are shown in Appendix~\ref{appendix:mixed-population}.
    }
\label{pop_MIX_A_B}
\end{figure*}

We present our results for the heterogeneous population Model $\mathrm{A}+\mathrm{B}$ in Figure~\ref{pop_MIX_A_B}. 
As previously and as expected, we see that the injected values are generally well recovered, with small offsets in the posterior depending upon the realisation of the catalogue and simulated detector noise.
The hyperparameters associated with the same physical parameters in the two components exhibit weak correlations.
Since the Model A component contributes $70\%$ to this mixed model, we observe that the hyperparameters of Model A, which are restricted to the \ac{MBH} mass and spin, are well constrained.
The hyperparameters of \ac{MBH} mass and spin: $x_c^{\mathrm{A}}$ and $\alpha_a^{\mathrm{A}}$, $\beta_a^{\mathrm{A}}$ are recovered with less than the fractional uncertainty of $2\%$ and $4\%$ respectively.
While the spin hyperparameters, $\mu_a^{\mathrm{B}}$ and $\sigma_a^{\mathrm{B}}$ of the Model B component are still recovered with sub-percent precision, the mass hyperparameters of the \ac{CO} are poorly constrained, evident by the broadened posteriors of the Model B component.
The branching fraction $w$ is recovered within $2\%$, demonstrating that the framework is capable of differentiating between the two distinct population models.

For Model $\mathrm{A^{(1)}}+\mathrm{A^{(2)}}$, we recover hyperposteriors when the subpopulations differ significantly in their underlying hyperparameter values, rather than in their parameterisation (Figure~\ref{pop_MIX_A_A} in Appendix~\ref{appendix:mixed-population}).
This confirms that the framework can successfully detect population diversity even within a single model type, provided that the subpopulations are sufficiently distinct.
The mass-related hyperparameters show broad posteriors and correlated regions between the mass hyperparameters of \ac{MBH} ($x_c^\mathrm{A^{(1/2)}}$) and \ac{CO} ($\gamma_\mu^\mathrm{A^{(1/2)}}$).
Similarly, for Model $\mathrm{B^{(1)}}+\mathrm{B^{(2)}}$ (Figure~\ref{pop_MIX_B_B} in Appendix~\ref{appendix:mixed-population}), mass hyperparameters are strongly correlated, and a larger uncertainty is associated with their estimation.
However, we also observe bimodality for the spin hyperparameters and branching fraction. 
Each mode in the recovered posterior corresponds to the injected extremal values of the hyperparameters.
Even though the mean spin hyperparameters ($\mu_a^\mathrm{B^{(1/2)}}$) are recovered to within $\sim 1\%$, the corresponding variances ($\sigma_a^\mathrm{B^{(1/2)}}$) exhibit substantially larger uncertainties of order $\sim 4\%$.
This increased uncertainty also propagates to the branching fraction $w$; this has a total uncertainty of $\sim60\%$, although the width of the single mode corresponds to an uncertainty of $\sim 2\%$.
Although the parameters remain well constrained, the dual peaks cluster around the true injected values, indicating that the framework correctly identifies two underlying populations rather than averaging them out.

The convergence behaviour remains consistent across all mixed model cases. 
Increasing the number of injected events yields tighter constraints and sharper identification of subpopulation features.
However, due to the presence of weak correlations between the hyperparameters associated with the same physical parameters, we expect the rate of improvement to deviate from $1/\sqrt{N_{\mathrm{det}}}$.
These results demonstrate that the inference framework remains robust for mixed populations and can resolve composite structure, even when the subpopulations share similar functional forms.

\subsection{Inference under model misspecification}
\label{case_3_inference}

\begin{figure*}
    \centering
    \begin{subfigure}[]{\includegraphics[width=8cm,height=10cm]{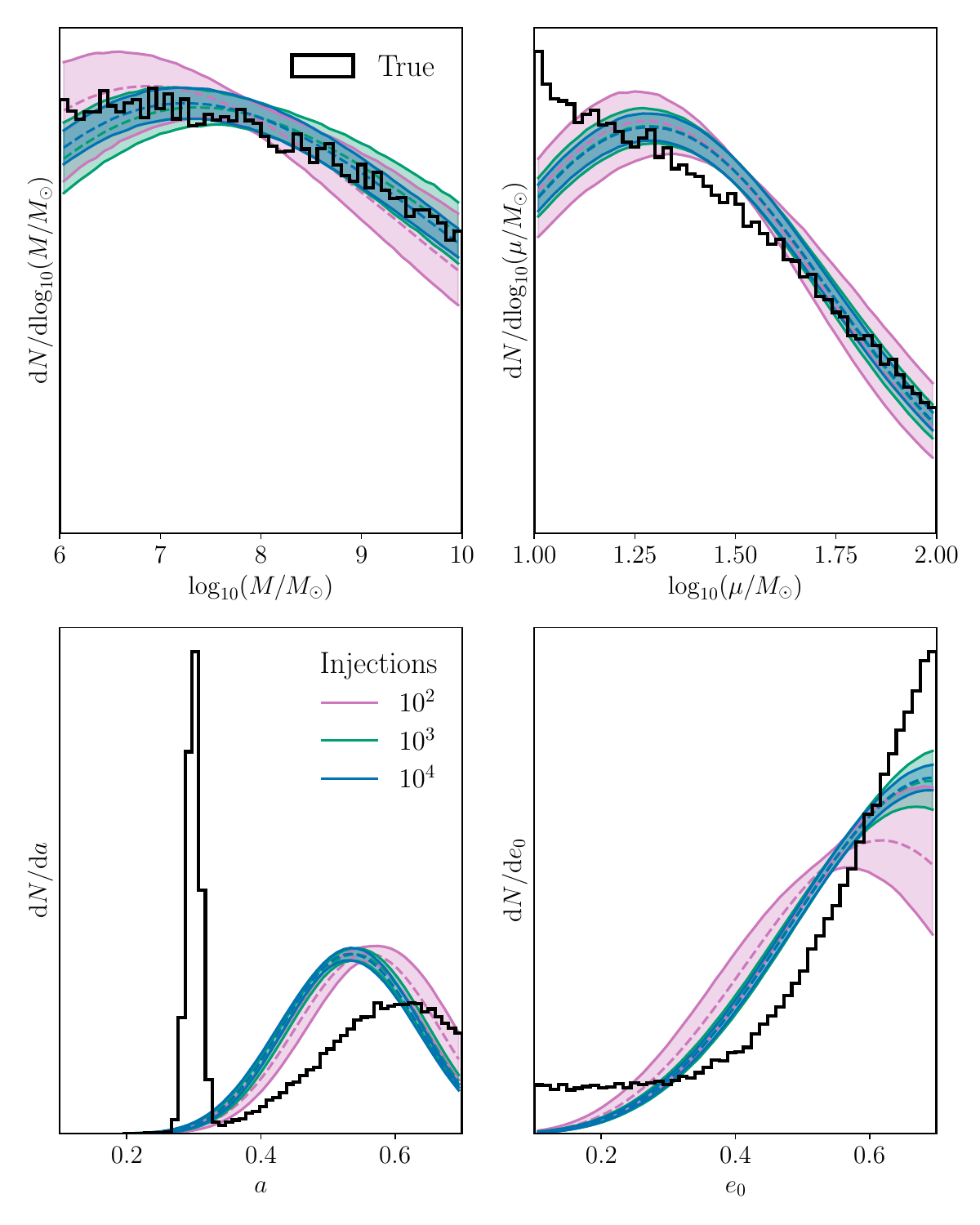}}\end{subfigure}
    \begin{subfigure}[]{\includegraphics[width=8cm,height=10cm]{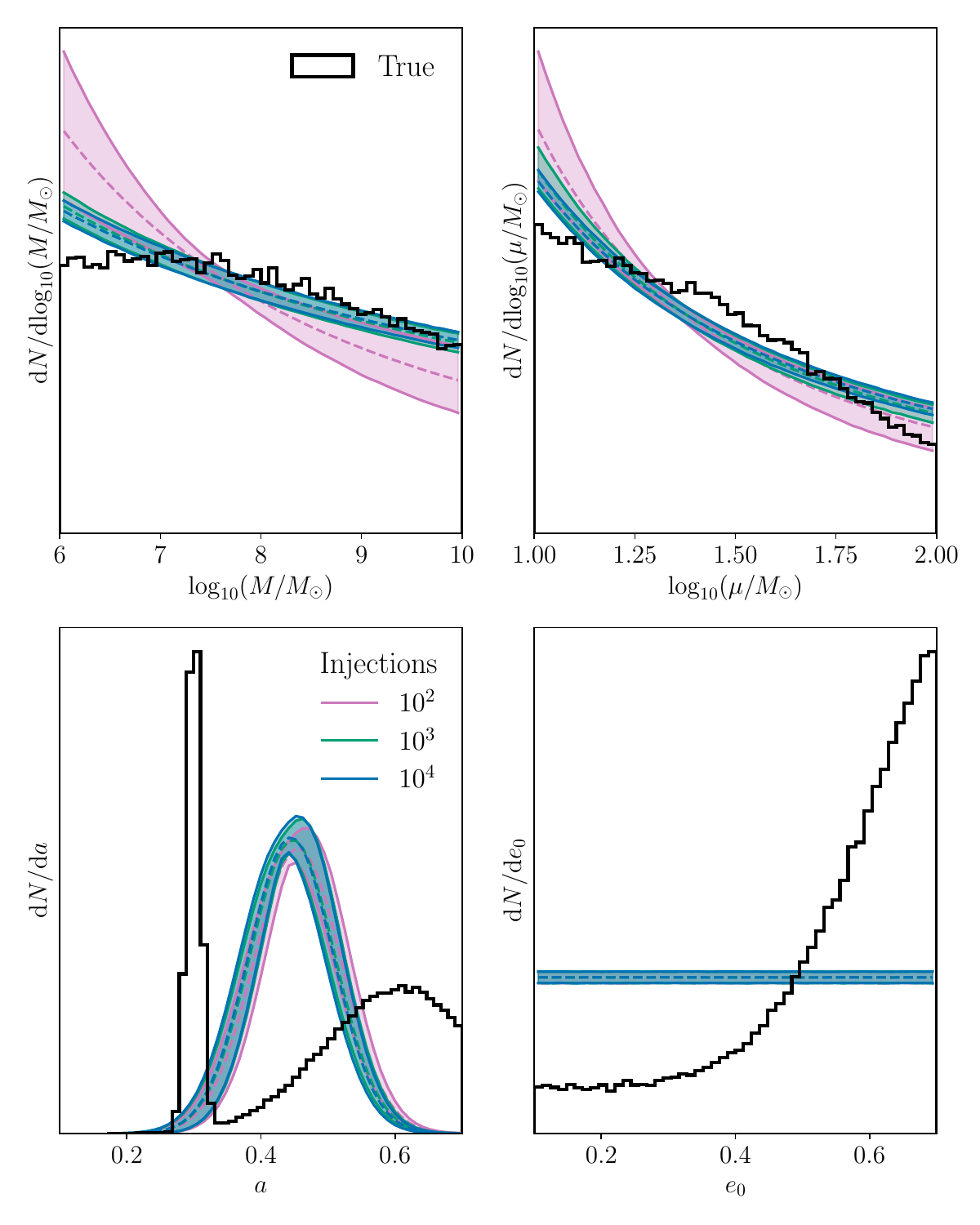}}\end{subfigure}
    \caption{\ac{PPD} plots for the population Model $\mathrm{A}+\mathrm{B}$ inferred with Model A (\textit{left}) and B (\textit{right}).
    The coloured solid curves denote the $90$th percentile and the dashed curves denote the mean.
    The colour scheme is consistent across panels: pink, green, and blue correspond to $10^2$, $10^3$, and $10^4$ \ac{EMRI} injections, respectively.
    The solid black curve represents the true distribution reconstructed from the injected hyperparameters.
    We display the differential number density $\mathrm{d}N/\mathrm{d}\boldsymbol{\theta}_i$ for $\boldsymbol{\theta}_i \in \{\log_{10}(M/M_\odot), \log_{10}(\mu/M_\odot), a, e_0\}$, with symbols defined in Table~\ref{tab:models}.
    }
    \label{population_description_plot_wrong_infer_MIX_A_B}
\end{figure*}

In this final experiment, we assess the robustness of the hierarchical inference framework to model misspecification.
This is an important consideration given that real astrophysical populations are unlikely to be perfectly represented by any single model.
We evaluate cases in which the inferred population model does not match the true underlying distribution.
Specifically, we infer Model A using Models B and $\mathrm{A}+\mathrm{B}$, and Model B using Models A and $\mathrm{A}+\mathrm{B}$.
We also extend this analysis by using Model A and B separately to recover the mixed $\mathrm{A}+\mathrm{B}$ population model.
This allows us to quantify how an incorrect functional form biases the recovered hyperparameters and to assess whether key population features remain identifiable despite model mismatch.

We present our analysis in Figure~\ref{population_description_plot_wrong_infer_MIX_A_B}, which highlights several systematic patterns that emerge under model mis-specification.
Since in Model $\mathrm{A}+\mathrm{B}$, the Model A subpopulation is the dominant component ($70\%$), we see that the mean of the reconstructed distribution closely matches the true distribution.
We observe that the \ac{MBH} mass, which follows a Schechter distribution, is well reconstructed compared to the power-law contribution of Model B, which is subdominant.
This trend is also similar for the \ac{CO} mass spectrum.
The skewness $\gamma_\mu$ can be adjusted to account for the shape of the mixed Model $\mathrm{A}+\mathrm{B}$ population.
However, we observe that the portion of the low-end \ac{CO} mass spectrum which is modelled by B is not well reconstructed.
A larger variance could potentially fit it, but the inferred distribution primarily reconstructs the dominant population.
The reconstruction of the \ac{MBH} and \ac{CO} distribution improves with an increase in the number of injections, closely matching the true value.

While the \ac{MBH} mass distribution is well reconstructed, this is not true for other distributions.
The spin of the \ac{MBH} and the eccentricity are poorly reconstructed.
Since we are inferring the whole population model with a single component, the inference framework tends to average out the extremities.
This is a behaviour we observe in the \ac{MBH} spin where the mean of the beta distribution is skewed towards the median of the distribution.
We also see that the reconstructed beta distribution is skewed towards the peak of the true Model A component, resulting in $ \beta_a < \alpha_a$, which corresponds to higher spins.
This skewness reflects the influence of our modelling choice: inferring a mixed population model using a single-component model.
A similar trend is observed for the \ac{CO} mass spectrum.
For eccentricities, we observe a similar feature, even though the averaging here is not immediately apparent.
The hyperparameters of the beta distribution $\alpha_{e_0}$ and $\beta_{e_0}$ are unable to fully recover the uniform part of the distribution.
Instead, they adjust to partly capture the subdominant population, while still remaining skewed towards high eccentricity, which is the dominant component.

When recovering the Model $\mathrm{A}+\mathrm{B}$ with Model B, we recover a similar trend.
For the \ac{MBH} mass distribution, the inferred power-law flattens to accommodate the strong high-mass contribution from Model A’s Schechter-like distribution (peaked $\sim10^7~M_{\odot}$).
This adjustment biases the recovered distribution toward higher masses, preventing the accurate reconstruction of the lower-mass component.
A similar trend is observed in the \ac{CO} mass spectrum, where the slope also shifts to favour the dominant low-mass spectrum.

The \ac{MBH} spin follows a similar trend when $\mathrm{A}+\mathrm{B}$ is modelled with A such that the estimated mean is around the median of the distribution.
Due to our uniform modelling of the eccentricity for Model B, it completely failed to capture the features present in the eccentricity distribution of Model $\mathrm{A}+\mathrm{B}$.

The reconstruction of Model A with $\mathrm{A}+\mathrm{B}$ yields a better reconstruction compared to reconstructing Model A with B alone, for a similar reason that A has a dominant contribution in the $\mathrm{A}+\mathrm{B}$ model.
Both the parameters \ac{MBH} and \ac{CO} mass spectra are well reconstructed; however, the lower and higher ends of the spectrum show broadening in the $90$th percentile lines, indicating the presence of a subdominant population.
For the \ac{MBH} spin distribution, the $90$th percentile lines show significant bumps and broadening near the mean of the subdominant B component.
In contrast, reconstruction using Model B, which has a completely different functional form, results in poor reconstruction. 
For the \ac{MBH} and \ac{CO} mass spectra, the true reconstructed distribution falls outside the $90$th percentile mass spectrum regions, showcasing the difficulty of using a power law to fit a distribution that possesses features.
Similarly, for the \ac{MBH} spin reconstruction, the estimated mean deviates by approximately $30\%$ from the injected mean while the eccentricity remains unconstrained due to its uniform underlying distribution.

Conversely, when we infer Model B from Model A and $\mathrm{A}+\mathrm{B}$, the reconstruction is poor across all parameters.
In both cases, the \ac{MBH} and \ac{CO} mass reconstructions attempt to capture the power-law structure in the data, whereas the inference for spin aims to capture the narrow peak in the spin distribution.
However, neither Model A nor Model $\mathrm{A}+\mathrm{B}$ successfully reconstructs a portion of the true spectrum.
Despite the overall poor performance, the inferred spin distribution exhibits a skew toward higher values, partially mimicking the behaviour observed in the data.
This behaviour is consistent across both models.
The inference compensates by pushing parameters toward their extreme values, attempting to mimic the narrower true distribution within the broader functional family of Model A.
This behaviour reflects the beta distribution's ability to reproduce the sharp central clustering of the truncated normal model.

These results demonstrate that population inference under model misspecification leads to systematic, model-dependent biases rather than purely statistical fluctuations.
The nature and magnitude of these biases depend on both the true underlying population and the functional form adopted for inference.
In particular, when the assumed function form lacks the flexibility to represent salient features of the true distribution, the inference compensates by shifting hyperparameters toward values that best approximate the dominant component of the population, often at the expense of subdominant features.
As a result, the recovered distributions tend to track the most prominent structures in the data while smoothing over or misrepresenting secondary, subdominant components.
We find recurring signatures in the \acp{PPD}, including broadened credible intervals and shifts in the reconstructed means relative to the injected population.
While such features do not, by themselves, uniquely diagnose model misspecification, their systematic appearance across different parameters and inference setups provides evidence of tension between the assumed model and the underlying population.
Posterior predictive checks, therefore, offer a useful consistency test for assessing whether an inferred model adequately captures the observed data \citep{LIGOScientific:2020kqk, Romero-Shaw:2022ctb}.
Even under incorrect assumptions, however, the inference retains sensitivity to physical parameters, suggesting that \ac{EMRI} population inference can still deliver meaningful astrophysical insight, provided that priors and model families are carefully chosen.

\section{Conclusion}
\label{sec:conclusion}

With their sub-percent level of constraints on \ac{MBH} and \ac{CO} masses, \ac{MBH} spin, and orbital eccentricity, \acp{EMRI} provide a promising opportunity to disentangle the \ac{MBH} formation mechanisms. 
We have investigated how a set of \ac{EMRI} observations can be used to infer the underlying astrophysical population. 
We use a Bayesian hierarchical inference framework that accounts for the selection biases, overcoming the associated computational obstacle by leveraging the \ac{ML} \texttt{poplar} emulator to obtain unbiased estimates of the population hyperparameters~\citep{Chapman-Bird:2022tvu, Singh:2025pzh}. 
We performed population inference using a combination of single-component (A and B) and mixed ($\mathrm{A^{(1)}}+\mathrm{A^{(2)}}$, $\mathrm{A}+\mathrm{B}$, $\mathrm{B^{(1)}}+\mathrm{B^{(2)}}$) population models for \ac{EMRI} sources to explore what \textit{LISA} could tell us about the \ac{EMRI} source population.

For single-component population models A and B, our population inferences reliably recover the true hyperparameters with increasing precision as the number of detections grows.
The constraining power is closely linked to the structure of the underlying distributions.
Sharply defined features, such as the peak of the \ac{MBH} mass spectrum in Model A (a Schechter distribution) or the narrow width of a Gaussian spin distribution, produce tight constraints. 
Broader distributions yield larger uncertainties. 
These results confirm that, in the absence of model mismatch, hierarchical inferences can extract population-level information from \ac{EMRI} catalogues.

Considering the plausible scenario that the \ac{EMRI} population comes from multiple channels, we also studied mixed populations.
Results demonstrate that population inferences can identify and characterise composite structure, even when subpopulations share the same general functional forms.
For heterogeneous mixtures ($\mathrm{A}+\mathrm{B}$), the inference can separate components with distinct astrophysical origins and parametrisations, recovering the branching fraction with percent–level accuracy for $10^2$ injections ($19$ detections).
For homogeneous mixtures ($\mathrm{A^{(1)}}+\mathrm{A^{(2)}}$, $\mathrm{B^{(1)}}+\mathrm{B^{(2)}}$), inferences can highlight internal diversity driven solely by contrasting hyperparameter values.
In these cases, correlations and partial degeneracies naturally emerge between subpopulation parameters, but the parametrised population inference framework nonetheless retains sensitivity to multimodality and avoids collapsing to a single averaged solution when $\mathcal{O}(>10^2)$ observations are available.

It is unlikely that a parametrised population model will precisely describe the true source population. 
Considering model misspecification, we find that systematic biases arise in predictable and physically interpretable ways.
Complex models can partially mimic simpler ones by tuning hyperparameters across a wide range, whereas simpler models fail to reconstruct the full shape of more structured populations.
The direction of these biases provides diagnostic signatures of functional mismatch, highlighting the importance of choosing flexible yet physically grounded population models.
Despite incorrect assumptions, key population trends often remain detectable, underscoring the resilience of hierarchical inference and the scientific value of \ac{EMRI} catalogues even when astrophysical modelling is imperfect.

An extension of this framework could be to perform model comparison between different \ac{EMRI} formation scenarios.
While we focus on parameter recovery and posterior predictive checks for a given population model, the same hierarchical formalism could be employed to compare how well a model is supported by the data.
Different formation channels, or astrophysical prescriptions for \ac{MBH} growth, can be treated as distinct population models, each characterised by its own set of hyperparameters, $\boldsymbol{\Lambda}$.
By computing the hyperevidence $\mathcal{Z}(\boldsymbol{d})$ for each model, we can calculate the \emph{Bayes factors} to quantify the models' relative support for the observed \ac{EMRI} catalogue~\citep{Kass:1995loi, Thrane:2018qnx}.
This provides another route for differentiating between alternative formation channels~\citep{Adams:2012qw}. 
Weighting the Bayes factors by prior probabilities for each model, would allow of posterior odds of different models to be evaluated.

In reality, we expect multiple formation channels to operate in parallel. 
Therefore, the task will not only be to distinguish between discrete models, but also to determine their relative mix. 
The possibility of a variable number of formation channels could be considered by extending our mixed models. 
For example, a population described solely by Model A can be considered as a \emph{nested} model of Model $\mathrm{A}+\mathrm{B}$ where the fraction from Model B is $0$~\citep{Trotta:2005ar}.  
For such nested models, Bayes factors can be computed efficiently using the Savage--Dickey density ratio~\citep{Dickey1971, Diciccio01091997}.
In this approach, the Bayes factor (Model A versus Model $\mathrm{A}+\mathrm{B}$) is obtained by comparing how strongly the ratio of the hyperposterior to the hyperprior at a specific value of the shared parameter that defines the simpler model. 
A key advantage of the Savage--Dickey density ratio is that it requires only the marginal hyperposterior of the extended model, which is already available as a by-product of parameter estimation, and does not rely on assumptions about the posterior distribution's shape.  
This is a computationally inexpensive route to Bayesian model comparison within the same inference framework, and is particularly well suited to situations where deviations from the simpler model are moderate.

Overall, our results collectively demonstrate the potential of \ac{EMRI} observations. 
We find that hierarchical population inference is resilient: while incorrect functional assumptions lead to unsurprising systematic biases, inference results nonetheless preserve qualitative sensitivity to key astrophysical trends.
This implies that, for future \ac{LISA} data analyses, even simplified or phenomenological models can extract meaningful constraints on \ac{EMRI} formation channels, provided that model selection is performed jointly with hierarchical inference.
The consistent recovery and strong constraints on spin--eccentricity correlations, along with the ability to measure relative mixture weights across various scenarios, highlight how future \ac{LISA} observations can extract meaningful astrophysical constraints on \ac{EMRI} formation channels, even when using simplified phenomenological models.

\begin{acknowledgments}
We thank Jonathan Gair for useful discussions. 
SS acknowledges support from the University of Glasgow.
CEAC-B acknowledges past support from STFC studentship 2446638 from grant number ST/V506692/1, and current support from UKSA grant UKRI971. 
CPLB acknowledges support from the UKSA grant UKRI972.
\end{acknowledgments}

\software{
\texttt{NumPy}~\citep{2020NumPy-Array},
\texttt{SciPy}~\citep{2020SciPy-NMeth},
\texttt{Matplotlib}~\citep{Hunter:2007},
\texttt{PyTorch}~\citep{2019arXiv191201703P},
\texttt{Nessai}~\citep{nessai},
\texttt{StableEMRIFisher}~\citep{kejriwal_2024_sef},
\texttt{poplar}~\citep{chapman_bird_2025_17897846},
\texttt{FastEMRIWaveforms}~\citep{chapman_bird_2025_15624459},
\texttt{fastlisaresponse}~\citep{Katz:2022yqe},
}

\section*{Data Availability}
All data used in this work are available on \href{https://doi.org/10.5281/zenodo.18197325}{Zenodo}, and the hierarchical inference analysis scripts are available on \href{https://github.com/SSingh087/MBHpopulation/tree/main/MBH_population_from_EMRI}{GitHub}.

\bibliography{EMRI-populations}{}
\bibliographystyle{aasjournalv7}

\appendix

\section{Inference plots for population Models $\mathrm{A^{(1)}}+\mathrm{A^{(2)}}$ and $\mathrm{B^{(1)}}+\mathrm{B^{(2)}}$}

\label{appendix:mixed-population}

\begin{figure*}[b]
    \centering
    \includegraphics[width=17cm, height=13cm]{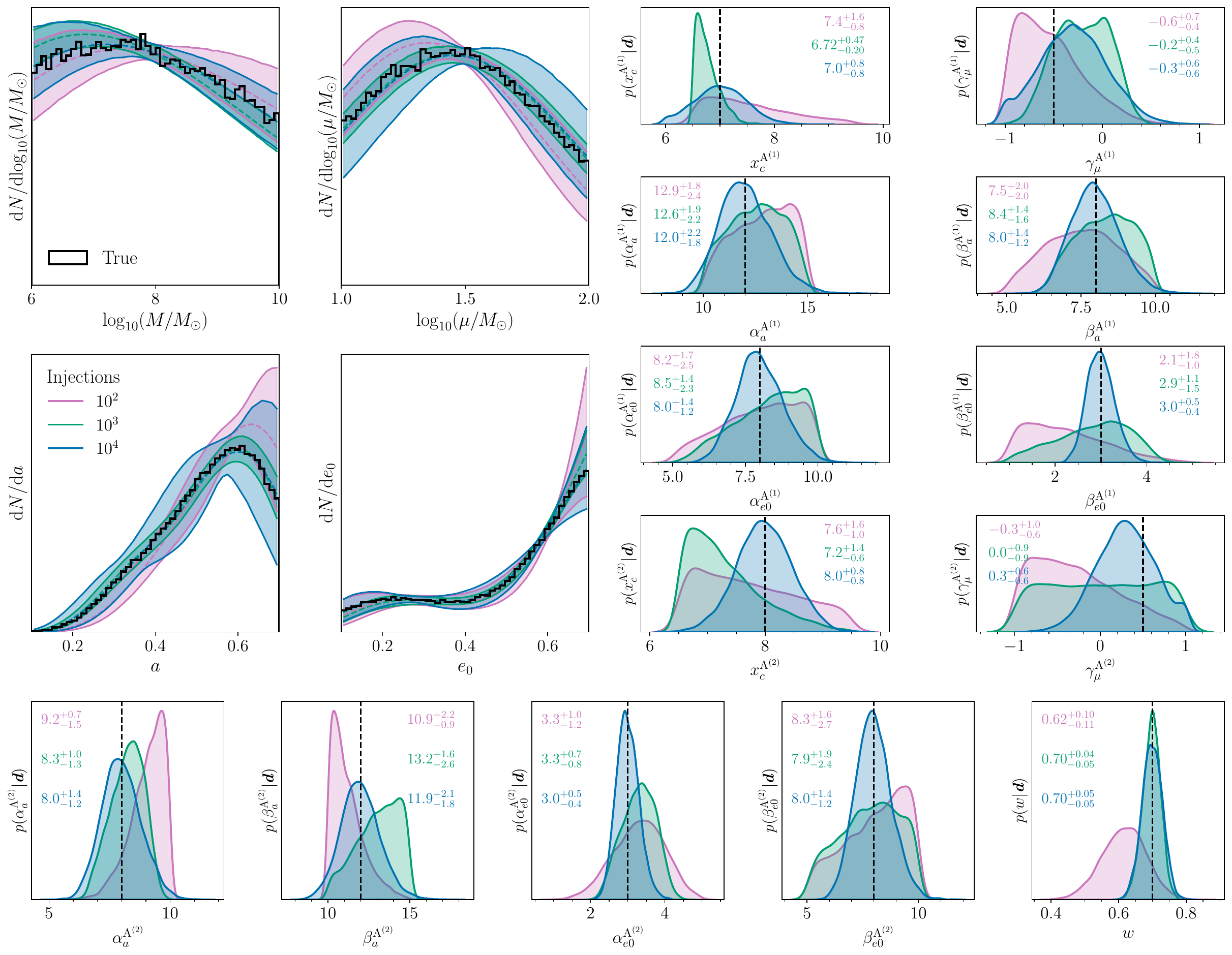}
    \caption{Posterior predictive distributions (\textit{upper left}) and one-dimensional hyperparameter posteriors (\textit{right} and \textit{bottom}) for Model $\mathrm{A^{(1)}}+\mathrm{A^{(2)}}$. 
    The plotting follows the same structure as Figure~\ref{pop_A}, where the only difference is the number of hyperparameters used to model the mixed population, where the number of detections is $19$, $196$ and $1988$, for $10^2$, $10^3$, and $10^4$ injections, respectively.
    }
\label{pop_MIX_A_A}
\end{figure*}

\begin{figure*}
    \centering
    \includegraphics[width=17cm, height=13cm]{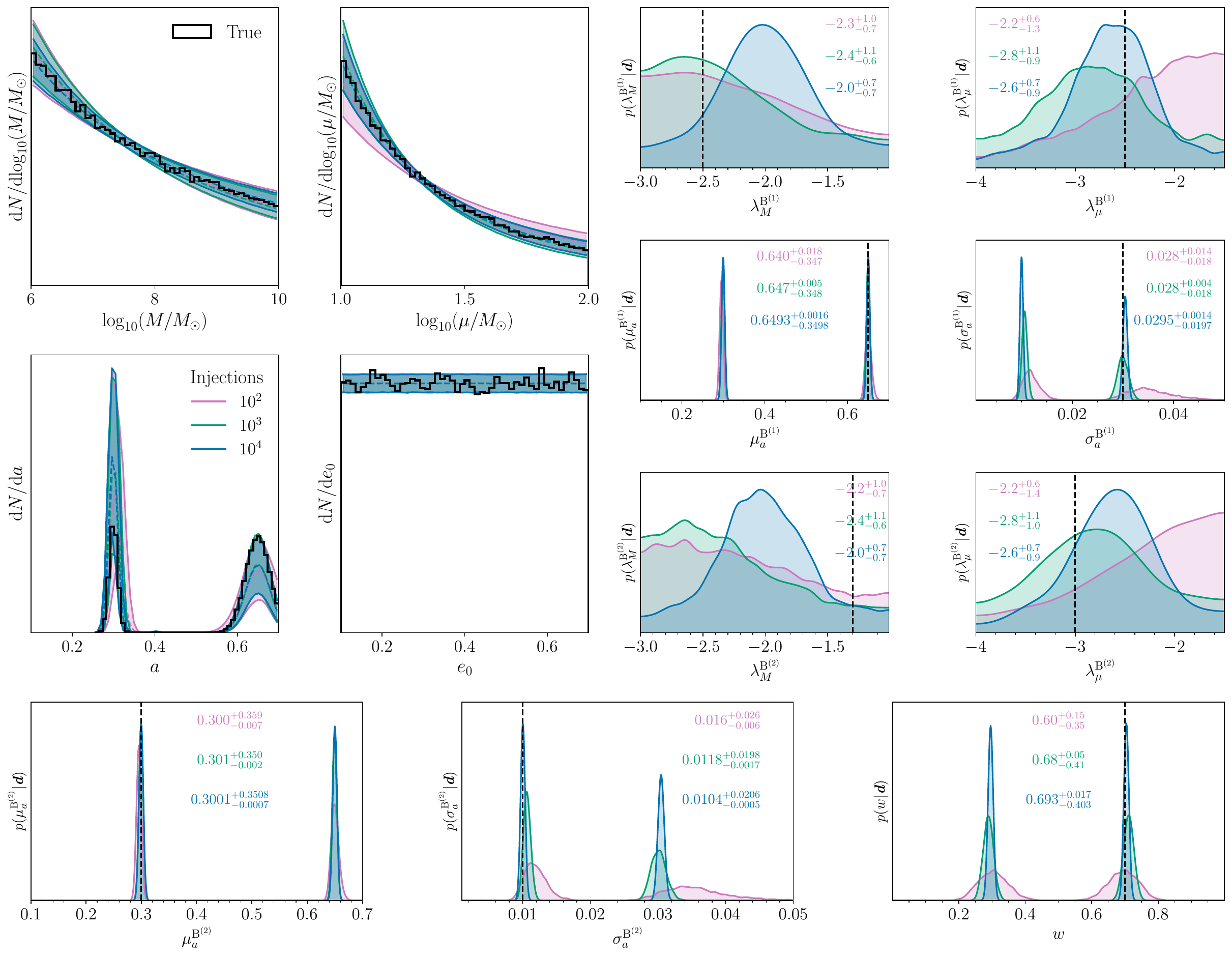}
    \caption{Posterior predictive distributions (\textit{upper left}) and one-dimensional hyperparameter posteriors (\textit{right} and \textit{bottom}) for Model $\mathrm{B^{(1)}}+\mathrm{B^{(2)}}$. 
    The plotting follows the same structure as Figure~\ref{pop_A}, where the only difference is the number of hyperparameters used to model the mixed population, where the number of detections is $21$, $212$ and $2134$ for $10^2$, $10^3$, and $10^4$ injections, respectively. 
    For bimodal distributions, the median lies in one of the modes, so the uncertainties are asymmetric, with the larger uncertainty reflecting the offset between the modes and the smaller reflecting the width of a single mode.
    }
\label{pop_MIX_B_B}
\end{figure*}

Figures~\ref{pop_MIX_A_A} and \ref{pop_MIX_B_B} present \acp{PPD} and one-dimensional hyperparameter posteriors for the mixed-population models $\mathrm{A^{(1)}}+\mathrm{A^{(2)}}$ and $\mathrm{B^{(1)}}+\mathrm{B^{(2)}}$, respectively.

For Model $\mathrm{A^{(1)}}+\mathrm{A^{(2)}}$, the hyperposteriors confirm that the inference pipeline successfully distinguishes between two subpopulations within the same model type.
While spin-related hyperparameters are inferred with sub-percent accuracy, mass-related hyperparameters exhibit broader posteriors and correlated regions, particularly between the \ac{MBH} mass parameters ($x_c^\mathrm{A^{(1/2)}}$) and \ac{CO} mass parameters ($\gamma_\mu^\mathrm{A^{(1/2)}}$).

For Model $\mathrm{B^{(1)}}+\mathrm{B^{(2)}}$, similar trends are observed: mass hyperparameters remain strongly correlated and less precisely constrained, while spin hyperparameters are recovered with high accuracy.
The spin posteriors display bimodality, with each mode corresponding to the injected extremal values of the hyperparameters.
This bimodality also propagates to the branching fraction $w$, which, despite being well constrained, shows two distinct peaks.
These dual modes indicate that the framework correctly identifies the presence of two underlying populations rather than averaging them out, demonstrating robustness in detecting population heterogeneity.

\end{document}